\PassOptionsToPackage{unicode}{hyperref}
\PassOptionsToPackage{hyphens}{url}
\PassOptionsToPackage{dvipsnames,svgnames,x11names}{xcolor}
\documentclass[
]{article}
\usepackage{amsmath,amssymb}
\usepackage{lmodern}
\usepackage{iftex}
\ifPDFTeX
  \usepackage[T1]{fontenc}
  \usepackage[utf8]{inputenc}
  \usepackage{textcomp} 
\else 
  \usepackage{unicode-math}
  \defaultfontfeatures{Scale=MatchLowercase}
  \defaultfontfeatures[\rmfamily]{Ligatures=TeX,Scale=1}
\fi
\IfFileExists{upquote.sty}{\usepackage{upquote}}{}
\IfFileExists{microtype.sty}{
  \usepackage[]{microtype}
  \UseMicrotypeSet[protrusion]{basicmath} 
}{}
\makeatletter
\@ifundefined{KOMAClassName}{
  \IfFileExists{parskip.sty}{%
    \usepackage{parskip}
  }{
    \setlength{\parindent}{0pt}
    \setlength{\parskip}{6pt plus 2pt minus 1pt}}
}{
  \KOMAoptions{parskip=half}}
\makeatother
\usepackage{xcolor}
\usepackage{graphicx}
\makeatletter
\def\maxwidth{\ifdim\Gin@nat@width>\linewidth\linewidth\else\Gin@nat@width\fi}
\def\maxheight{\ifdim\Gin@nat@height>\textheight\textheight\else\Gin@nat@height\fi}
\makeatother
\setkeys{Gin}{width=\maxwidth,height=\maxheight,keepaspectratio}
\makeatletter
\def\fps@figure{htbp}
\makeatother
\providecommand{\tightlist}{%
  \setlength{\itemsep}{0pt}\setlength{\parskip}{0pt}}
\newlength{\cslhangindent}
\newlength{\csllabelwidth}
\newlength{\cslentryspacingunit} 
\newenvironment{CSLReferences}[2] 
 {
  \setlength{\parindent}{0pt}
  \ifodd #1
  \let\oldpar\par
  \def\par{\hangindent=\cslhangindent\oldpar}
  \fi
  \setlength{\parskip}{#2\cslentryspacingunit}
 }%
 {}
\usepackage{calc}

\ifLuaTeX
\usepackage[bidi=basic]{babel}
\else
\usepackage[bidi=default]{babel}
\fi
\babelprovide[main,import]{american}

\def\languageshorthands#1{}
\ifLuaTeX
  \usepackage{selnolig}  
\fi
\IfFileExists{bookmark.sty}{\usepackage{bookmark}}{\usepackage{hyperref}}
\IfFileExists{xurl.sty}{\usepackage{xurl}}{} 
\hypersetup{
  pdftitle={KerrGeoPy: A Python Package for Computing Timelike Geodesics
in Kerr Spacetime},
  pdfauthor={Seyong Park, Zachary Nasipak},
  pdflang={en-US},
  colorlinks=true,
  linkcolor={Maroon},
  filecolor={Maroon},
  citecolor={Blue},
  urlcolor={Blue},
  pdfcreator={LaTeX via pandoc}}

\title{\texttt{KerrGeoPy}: A Python Package for Computing Timelike
Geodesics in Kerr Spacetime}

\usepackage[affil-it]{authblk}
\usepackage{orcidlink}
\author[1,2%
  ]{Seyong Park%
    \,\orcidlink{0009-0002-1152-9324}\,%
    }
\author[1%
  ]{Zachary Nasipak%
    \,\orcidlink{0000-0002-5109-9704}\,%
    }

\affil[1]{NASA Goddard Space Flight Center, Greenbelt, MD, USA}
\affil[2]{University of Maryland, College Park, MD, USA}
\date{15 December 2023}

\begin{document}
\maketitle

\hypertarget{summary}{%
\section{Summary}\label{summary}}

In general relativity, the motion of a free-falling test particle in a
curved spacetime is described by a timelike geodesic - the minimal path
between two points in space. Intuitively, geodesics are analogous to
straight line paths in flat spacetime. The timelike geodesics of Kerr
spacetime are of particular interest in the field of black hole
perturbation theory because they describe the zeroth-order motion of a
small object moving through the background spacetime of a much more
massive spinning black hole. For this reason, computing geodesics is an
important step in modeling the gravitational radiation emitted by an
extreme-mass-ratio inspiral (EMRI) - an astrophysical binary in which a
stellar mass compact object, such as a neutron star or black hole (with
mass \(10^1 - 10^2 M_\odot\)), spirals into a massive black hole (with
mass \(10^4 - 10^7 M_\odot\)).

Kerr spacetime has several nice properties which simplify the problem of
computing geodesics. Since it has both time-translation symmetry and
rotational symmetry, energy and (the \(z\)-component of) angular
momentum are conserved quantities. It is also equipped with a higher
order symmetry which gives rise to a third constant of motion called the
Carter constant. These three constants of motion, along with the spin of
the black hole, uniquely define a geodesic up to initial conditions
(\protect\hyperlink{ref-schmidt}{Schmidt, 2002}). Alternatively,
geodesics can be identified using a suitably generalized version of the
parameters used to define a Keplerian orbit (eccentricity, semi-latus
rectum, and inclination angle). Bound geodesics also possess fundamental
frequencies since their radial, azimuthal, and polar motions are
periodic.

\texttt{KerrGeoPy} is a Python package which computes both stable and
plunging timelike geodesics in Kerr spacetime using the analytic
solutions to the geodesic equation derived in Fujita \& Hikida
(\protect\hyperlink{ref-fujita}{2009}) and Dyson \& Meent
(\protect\hyperlink{ref-dyson}{2023}). It mirrors and builds upon much
of the functionality of the \texttt{KerrGeodesics}
(\protect\hyperlink{ref-kerrgeodesics}{Warburton et al., 2023})
Mathematica library. Geodesic solutions are written in terms of Legendre
elliptic integrals, which are evaluated using \texttt{SciPy}. Users can
construct a geodesic by providing the initial position and
four-velocity, or by providing either the constants of motion or the
Keplerian parameters described above.

\texttt{KerrGeoPy} provides methods for computing the four-velocity,
fundamental frequencies, and constants of motion associated with a given
geodesic and also implements the algorithm described in Stein \&
Warburton (\protect\hyperlink{ref-stein}{2020}) for finding the location
of the last stable orbit, known as the separatrix. The package also
includes several methods for visualizing and animating geodesics.

\texttt{KerrGeoPy} is a part of the \href{https://bhptoolkit.org}{Black
Hole Perturbation Toolkit}. The source code is hosted on
\href{https://github.com/BlackHolePerturbationToolkit/KerrGeoPy}{Github}
and the package is distributed through both
\href{https://pypi.org/project/kerrgeopy/}{PyPI} and
\href{https://anaconda.org/conda-forge/kerrgeopy}{conda-forge}.
Automated unit tests are run using
\href{https://github.com/BlackHolePerturbationToolkit/KerrGeoPy/actions/workflows/tests.yml}{Github
Actions} and comprehensive documentation is available on
\href{https://kerrgeopy.readthedocs.io/}{Read the Docs}.

\begin{figure}
\centering
\includegraphics{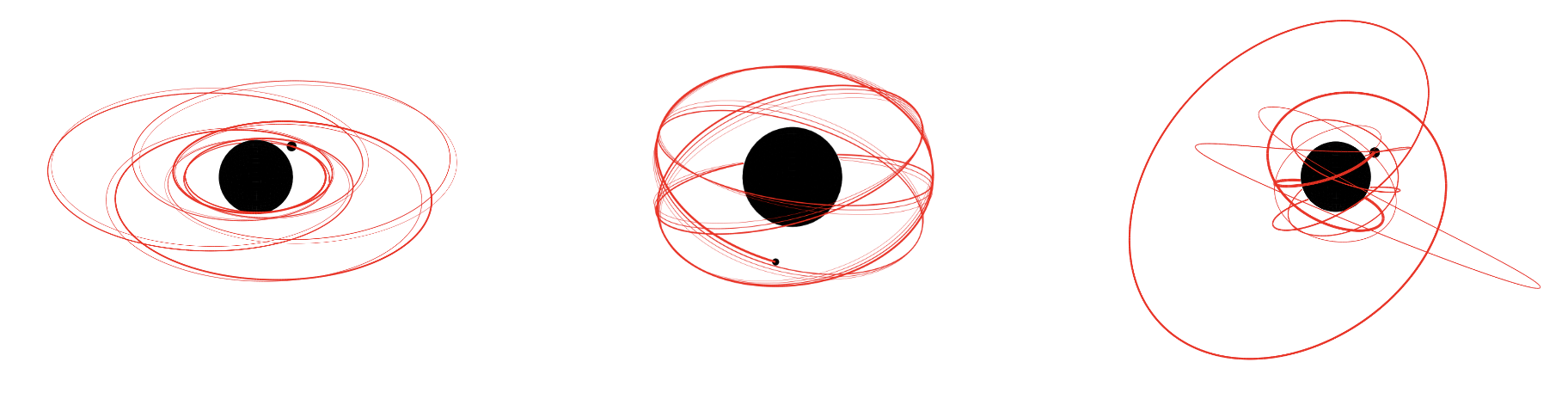}
\caption{Example of an equatorial (left), spherical (center) and generic
(right) orbit computed by \texttt{KerrGeoPy}}
\end{figure}

\hypertarget{statement-of-need}{%
\section{Statement of Need}\label{statement-of-need}}

EMRIs are expected to be a major source observable by the Laser
Interferometer Space Antenna (LISA), a future space-based gravitational
wave observatory consisting of a triangular constellation of three
satellites in orbit around the sun. LISA is an ESA-led mission with
significant contributions from NASA which is set to launch in the 2030s.
It will complement existing ground-based detectors by opening up the
millihertz band of the gravitational wave spectrum
(\protect\hyperlink{ref-lisa}{Thorpe et al., 2019}). Because sources in
this band evolve more slowly over time and remain observable for a
period of days to years, LISA is expected to detect many overlapping
signals at all times. Thus, accurate waveform models are needed in order
to identify gravitational wave sources and perform parameter estimation
- the process of approximating characteristics of a source.

For most LISA sources, well-developed waveform models based on either
numerical relativity or post-Newtonian theory already exist. However,
EMRIs are instead more naturally described by black hole perturbation
theory, and the EMRI waveform models that currently exist are
underdeveloped compared to other LISA sources. In a perturbation theory
model, the orbital trajectory is assumed to be a geodesic at leading
order. Higher-order corrections are then computed by introducing the
gravitational field of the inspiraling object as a perturbation to the
background spacetime of the massive black hole, expanded in powers of
the mass ratio.

To meet the accuracy requirements for LISA parameter estimation, EMRI
waveform models must include both first- and second-order corrections to
the orbital trajectory. However, to date, second-order corrections are
only available for the most simple systems, quasi-circular inspirals in
Schwarzschild (\protect\hyperlink{ref-emri}{Berry et al., 2019}).
Open-source tools can aid in rapidly expanding EMRI models to more
complicated orbits in Kerr spacetime, but at the moment many tools for
modeling EMRIs are only available in Mathematica, which is an expensive
and proprietary piece of software. \texttt{KerrGeoPy} is intended to
support future development of higher-order waveform models in
preparation for LISA by providing a free alternative to the existing
\texttt{KerrGeodesics} Mathematica library for other researchers to
build on in their own projects.

Although other Python packages
(\protect\hyperlink{ref-kerrgeodesicgw}{Gourgoulhon et al., 2019}) with
similar functionality do exist, they mostly rely on numerical
integration to compute geodesics. The analytic solutions used by
\texttt{KerrGeoPy} have two main advantages over this approach. First,
they can be much more numerically stable over long time periods and can
be quickly evaluated at any point in time. This is essential for EMRI
models, which typically require taking long time-averages over the
geodesic motion. Second, they produce several useful intermediate terms
which are not calculated by other packages. Therefore,
\texttt{KerrGeoPy}, with its analytic solutions and various orbital
parametrizations, is specifically tuned to support perturbative models
of binary black holes and their gravitational waves.

\newpage

\hypertarget{software-citations}{%
\section{Software Citations}\label{software-citations}}

\texttt{KerrGeoPy} has the following dependencies:

\begin{itemize}
\tightlist
\item
  \texttt{NumPy} (\protect\hyperlink{ref-numpy}{Harris et al., 2020})
\item
  \texttt{SciPy} (\protect\hyperlink{ref-scipy}{Virtanen et al., 2020})
\item
  \texttt{Matplotlib} (\protect\hyperlink{ref-matplotlib}{Hunter, 2007})
\item
  \texttt{tqdm} (\protect\hyperlink{ref-tqdm}{Costa-Luis et al., 2023})
\end{itemize}

\hypertarget{acknowledgements}{%
\section{Acknowledgements}\label{acknowledgements}}

We would like to thank Niels Warburton and Barry Wardell for their
assistance in releasing \texttt{KerrGeoPy} as part of the Black Hole
Perturbation Toolkit. SP acknowledges support through NASA's Office of
STEM Engagement, while ZN acknowledges support by an appointment to the
NASA Postdoctoral Program at the NASA Goddard Space Flight Center,
administered by Oak Ridge Associated Universities under contract with
NASA.

\hypertarget{references}{%
\section*{References}\label{references}}
\addcontentsline{toc}{section}{References}

\hypertarget{refs}{}
\begin{CSLReferences}{1}{0}
\leavevmode\vadjust pre{\hypertarget{ref-emri}{}}%
Berry, C., Hughes, S., Sopuerta, C., Chua, A., Heffernan, A.,
Holley-Bockelmann, K., Mihaylov, D., Miller, C., \& Sesana, A. (2019).
The unique potential of extreme mass-ratio inspirals for
gravitational-wave astronomy. \emph{Bulletin of the American
Astronomical Society}, \emph{51}, 42.
\url{https://doi.org/10.48550/arXiv.1903.03686}

\leavevmode\vadjust pre{\hypertarget{ref-tqdm}{}}%
Costa-Luis, C. da, Larroque, S. K., Altendorf, K., Mary, H.,
richardsheridan, Korobov, M., Yorav-Raphael, N., Ivanov, I., Bargull,
M., Rodrigues, N., Chen, G., Lee, A., Newey, C., CrazyPython, JC,
Zugnoni, M., Pagel, M. D., mjstevens777, Dektyarev, M., \ldots{} Boyle,
M. (2023). \emph{{tqdm: A fast, Extensible Progress Bar for Python and
CLI}} (Version v4.66.1). Zenodo.
\url{https://doi.org/10.5281/zenodo.8233425}

\leavevmode\vadjust pre{\hypertarget{ref-dyson}{}}%
Dyson, C., \& Meent, M. van de. (2023). Kerr-fully diving into the
abyss: Analytic solutions to plunging geodesics in {Kerr}.
\emph{Classical and Quantum Gravity}, \emph{40}, 195026.
\url{https://doi.org/10.1088/1361-6382/acf552}

\leavevmode\vadjust pre{\hypertarget{ref-fujita}{}}%
Fujita, R., \& Hikida, W. (2009). Analytical solutions of bound timelike
geodesic orbits in {Kerr} spacetime. \emph{Classical and Quantum
Gravity}, \emph{26}, 135002.
\url{https://doi.org/10.1088/0264-9381/26/13/135002}

\leavevmode\vadjust pre{\hypertarget{ref-kerrgeodesicgw}{}}%
Gourgoulhon, E., Le Tiec, A., Vincent, F. H., \& Warburton, N. (2019).
Gravitational waves from bodies orbiting the {Galactic} center black
hole and their detectability by {LISA}. \emph{Astronomy and
Astrophysics}, \emph{627}, A92.
\url{https://doi.org/10.1051/0004-6361/201935406}

\leavevmode\vadjust pre{\hypertarget{ref-numpy}{}}%
Harris, C. R., Millman, K. J., Walt, S. J. van der, Gommers, R.,
Virtanen, P., Cournapeau, D., Wieser, E., Taylor, J., Berg, S., Smith,
N. J., Kern, R., Picus, M., Hoyer, S., Kerkwijk, M. H. van, Brett, M.,
Haldane, A., Río, J. F. del, Wiebe, M., Peterson, P., \ldots{} Oliphant,
T. E. (2020). Array programming with {NumPy}. \emph{Nature},
\emph{585}(7825), 357--362.
\url{https://doi.org/10.1038/s41586-020-2649-2}

\leavevmode\vadjust pre{\hypertarget{ref-matplotlib}{}}%
Hunter, J. D. (2007). Matplotlib: {A} {2D} graphics environment.
\emph{Computing in Science \& Engineering}, \emph{9}(3), 90--95.
\url{https://doi.org/10.1109/MCSE.2007.55}

\leavevmode\vadjust pre{\hypertarget{ref-schmidt}{}}%
Schmidt, W. (2002). {Celestial mechanics in Kerr spacetime}.
\emph{Classical and Quantum Gravity}, \emph{19}(10), 2743--2764.
\url{https://doi.org/10.1088/0264-9381/19/10/314}

\leavevmode\vadjust pre{\hypertarget{ref-stein}{}}%
Stein, L. C., \& Warburton, N. (2020). Location of the last stable orbit
in {Kerr} spacetime. \emph{Physical Review D}, \emph{101}, 064007.
\url{https://doi.org/10.1103/PhysRevD.101.064007}

\leavevmode\vadjust pre{\hypertarget{ref-lisa}{}}%
Thorpe, J. I., Ziemer, J., Thorpe, I., Livas, J., Conklin, J. W.,
Caldwell, R., Berti, E., McWilliams, S. T., Stebbins, R., Shoemaker, D.,
Ferrara, E. C., Larson, S. L., Shoemaker, D., Key, J. S., Vallisneri,
M., Eracleous, M., Schnittman, J., Kamai, B., Camp, J., \ldots{} Wass,
P. (2019). \emph{The {Laser} {Interferometer} {Space} {Antenna}:
{Unveiling} the {Millihertz} {Gravitational} {Wave} {Sky}}. \emph{51},
77. \url{https://doi.org/10.48550/arXiv.1907.06482}

\leavevmode\vadjust pre{\hypertarget{ref-scipy}{}}%
Virtanen, P., Gommers, R., Oliphant, T. E., Haberland, M., Reddy, T.,
Cournapeau, D., Burovski, E., Peterson, P., Weckesser, W., Bright, J.,
Walt, S. J. van der, Brett, M., Wilson, J., Millman, K. J., Mayorov, N.,
Nelson, A. R. J., Jones, E., Kern, R., Larson, E., \ldots{} SciPy 1.0
Contributors. (2020). {SciPy} 1.0: {Fundamental} {Algorithms} for
{Scientific} {Computing} in {Python}. \emph{Nature Methods}, \emph{17},
261--272. \url{https://doi.org/10.1038/s41592-019-0686-2}

\leavevmode\vadjust pre{\hypertarget{ref-kerrgeodesics}{}}%
Warburton, N., Wardell, B., Long, O., Upton, S., Lynch, P., Nasipak, Z.,
\& Stein, L. C. (2023). \emph{{KerrGeodesics}}. Zenodo.
\url{https://doi.org/10.5281/zenodo.8108265}

\end{CSLReferences}

\end{document}